\documentclass[fleqn,usenatbib]{mnras}

\usepackage{newtxtext,newtxmath}
\usepackage[T1]{fontenc}

\DeclareRobustCommand{\VAN}[3]{#2}
\let\VANthebibliography\thebibliography
\def\thebibliography{\DeclareRobustCommand{\VAN}[3]{##3}\VANthebibliography}

\usepackage{graphicx}	% Including figure files
\usepackage{amsmath}	% Advanced maths commands

\usepackage{tabto} %REMOVE WHEN LISTS ARE GONE

\title[Spectrum de-noising]{De-noising of galaxy optical spectra with autoencoders}

\author[M.~Scourfield et al.]{M.~Scourfield$^{1}$\thanks{E-mail: matt.scourfield.18@ucl.ac.uk},
A.~Saintonge$^{1}$\thanks{E-mail: a.saintonge@ucl.ac.uk},
D. de Mijolla$^{1}$,
S.~Viti$^{2,1}$
\\
$^{1}$Department of Physics and Astronomy, University College London, Gower St., London, WC1E 6BT, UK \\
$^{2}$Leiden Observatory, Leiden University, PO Box 9513, 2300 RA Leiden, The Netherlands
}

\date{Accepted XXX. Received YYY; in original form ZZZ}

\pubyear{2021}

\begin{document}
\label{firstpage}
\pagerange{\pageref{firstpage}--\pageref{lastpage}}
\maketitle

\begin{abstract}
%Context (optional - not recomended):
%Aims:
Optical spectra contain a wealth of information about the physical properties and formation histories of galaxies. Often though, spectra are too noisy for this information to be accurately retrieved.  
%Methods:
In this study, we explore how machine learning methods can be used to de-noise spectra and increase the amount of information we can gain without having to turn to sample averaging methods such as spectral stacking. 
Using machine learning methods trained on noise-added spectra - SDSS spectra with Gaussian noise added - we investigate methods of maximising the information we can gain from these spectra, in particular from emission lines, such that more detailed analysis can be performed. 
%Results:
We produce a variational autoencoder (VAE) model, and apply it on a sample of noise-added spectra. Compared to the flux measured in the original SDSS spectra, the model values are accurate within 0.3-0.5 dex, depending on the specific spectral line and S/N. Overall, the VAE performs better than a principle component analysis (PCA) method, in terms of reconstruction loss and accuracy of the recovered line fluxes. 
%Conclusion (optional - recomended):
To demonstrate the applicability and usefulness of the method in the context of large optical spectroscopy surveys, we simulate a population of spectra with noise similar to that in galaxies at $z=0.1$ observed by the Dark Energy Spectroscopic Instrument (DESI). We show that we can recover the shape and scatter of the MZR in this "DESI-like" sample, in a way that is not possible without the VAE-assisted de-noising. 
\end{abstract}
\begin{keywords}
methods: data analysis processing -- galaxies: general
\end{keywords}

%%%%%%%%%%%%%%%%% BODY OF PAPER %%%%%%%%%%%%%%%%%%
\section{Introduction}

Spectroscopy is a particularly powerful tool to retrieve fundamental information about the contents of galaxies and their formation histories. Some quantities can be directly retrieved from spectral measurements (e.g. the total mass of the neutral interstellar medium from the H{\sc I} 21cm integrated line flux), others require complex modeling of continuum and/or line features (e.g. star formation histories). Especially in the latter case, having access to quality spectra with high resolution and sensitivity to both continuum and line emission is a requirement. In this paper, we focus on optical spectroscopy in the context of galaxy evolution studies, and explore methods to retrieve a maximum amount of information from spectra even when these sensitivity requirements are not met. 

Astronomy is experiencing a "data volume" revolution, and optical spectroscopy is no exception. For example, the Dark Energy Spectroscopic Instrument \citep[DESI;][]{DESI_part1, DESI_EDR} will produce spectra for in excess of 30 million galaxies, an order of magnitude more than was achieved by the Sloan Digital Sky Survey \citep[SDSS;][]{york00}. Such a data volume opens up new parameter space for galaxy evolution studies, but to make the most of these data we need to overcome the challenge that the depth of the spectra, while sufficient to measure redshifts, often does not allow for the accurate derivation of quantities such as metallicities, star formation rates/histories and stellar masses/kinematics. 

There are several methods to increase the S/N of spectra to overcome such challenges. Spectral stacking is currently the most commonly-used of these methods. By averaging the spectra, we can boost the signal and make measurements that are representative of the sub-population of objects stacked, even when individual objects do not have detectable emission. While a powerful tool, stacking also has its limitations, for example it does not return any information about the scatter in the underlying population and the results are entirely dependent on the a priori choices of what galaxies to combine into different stacks \citep[see][for further details]{ARAAreview}. 

Alternatively, methods which are agnostic to the physics of galaxies can also be used. One example of this is Principal Component Analysis \citep[PCA, ][]{PCA_original, PCA_review}, a type of dimensional reduction algorithm. In PCA, components are found which can be linearly combined to approximately reproduce an input. For example, \citet{PCA_SDSS_lin} find three components to be sufficient to reproduce most galaxy types found in the SDSS spectroscopic sample to within a 10\% error; increasing to ten components allows the modeling of rare populations such as galaxies with extreme emission lines.  While the linear nature of PCA lends itself well to interpretation of the components, it also limits the use of the method in reproducing data with non-linear components. For example, \citet{PCA_SDSSS_nonlin} apply PCA to the SDSS spectra of quasars and find that $\sim100$ components are required to accurately reproduce their sample out to $z \sim 5$, due to the non-linearity of the problem.

Machine learning (ML) techniques are also an example of these agnostic methods, and are increasingly used due to their flexibility and ability to process large quantities of data. ML has been used for spectral classification \citep{NN_dwarf, CNN_merger, ensemble_class}, artifact and anomaly detection \citep{GAN_anomaly, CNN_deepShadows},  parameter extraction \citep{NetZ, CNN_SFH, ML_Lya} and even as a method of going from image to spectra \citep{CNN_img2spec, CNN_img2spec_ap}. In particular,  Generative Adversarial Networks (GANs), a type of neural network, are commonly used as a method of de-noising and enhancing for astronomical images \citep{GAN_denoise, GAN_weak}. Another type of neural network commonly used for de-noising are Autoencoders \citep[AEs,][]{AE_denoise}, and have previously been applied to various areas of Astronomy, such as galaxy spectral energy distribution classification \citep{AE_SED}, gravitational waves analysis \citep{AE_GW_denoise, AE_GW_gen}, cosmological parameter constraint \citep{AE_cosmo}, deep cluster detection \citep{AE_cluster} and chemical modelling \citep{chemulator}.

In this paper, we investigate the use of autoencoders to perform spectral de-noising, and enhancing the amount of information we are able to retrieve from low S/N galaxy spectra. In section \ref{sec:data} we present the data used in this paper, and the various models we use are introduced in section \ref{sec:methods}. We then apply these models to SDSS spectra and discuss in Section \ref{sec:results} how they perform in accurately recovering emission line fluxes. As a proof-of-concept, we show how the autoencoder de-noising can enable the study of the mass-metallicity in a DESI-like survey in Section \ref{sec:metallicity} before presenting our conclusions in \ref{sec:conclusions}.

\section{Data}
\label{sec:data}

We use spectra from SDSS Data Release 7 \citep[DR7,][]{SDSS_DR7} in order to train our model. A lower redshift cutoff of $z = 0.024$ is chosen to ensure coverage of the [OII]3727\AA{} doublet, and an upper cutoff of $z = 0.05$, to prevent the sample being dominated by massive galaxies and to ensure good SNR of both emission lines and continuum. This leaves us with a total of 86,908 galaxies whose spectra can be used to construct our sample from.

\begin{table}
    \centering
    \begin{tabular}{c c c}
        Subclassification & Candidates & Sample \\
        \hline\hline
        Emission Line Galaxy    & 50948 &  4522 \\
        Absorption Galaxy       & 20771 &  4003 \\
        Quiescent Galaxy        & 12204 &  1175 \\
        Narrow-Line AGN         &  2968 &   297 \\
        Broad-Line AGN          &    17 &     3 \\
        \hline
        Total                   & 86908 & 10000
    \end{tabular}
    \caption{Breakdown of spectral subclassifications.}
    \label{tab:class_break}
\end{table}

These spectra each have a subclassification assigned to them from the SDSS and astroML \citep{SDSS_que} pipelines, which are the same as those presented in \cite{astroML_subclasses}: absorption galaxy, quiescent galaxy, emission line galaxy, narrow-line AGN and broad-line AGN. These subclassifications are determined by first using the output of the SDSS spectroscopic pipeline, splitting the spectra into galaxies and QSOs (here called broad-line AGN). The galaxy spectra are then further split; those with both H$\alpha$ and H$\beta$ emission lines with SNR$>3$ are classified as either narrow-line AGN or emission line galaxies based on their position in the BPT diagram according to the criterion of \citet{astroML_split}, and the remaining spectra are classified as either absorption-line galaxies if they have Balmer absorption lines detected with SNR$>3$, or quiescent galaxies if they show neither significant Balmer line emission or absorption. The breakdown of galaxy subclassifications within the candidates is shown in table \ref{tab:class_break}.

The properties of these candidate spectra are not uniformly distributed, meaning that randomly constructing our sample from the candidate spectra would result in models that are overfit to the most dense regions of the parameter space. As such, rather than randomly constructing our sample from the candidate spectra we instead use the derived galaxy properties from the MPA-JHU emission line analysis for SDSS DR7 to split spectra based on H$\alpha$ flux. We use this quantity as in this paper we are most concerned with reproducing the emission lines, rather than continuum features. This will be explored in a further study. We use 10 bins logarithmically spaced from ~$10^{0.2}-10^{3.4}$ 1e-17 erg/s/cm$^2$, populating each bin with 1,000 galaxies randomly selected out of all of those with H$\alpha$ flux in the relevant range. This gives us a final sample of 10,000 galaxies, of which 8,000 are used during model training and validation, and the remaining 2,000 for testing performance. During training of neural network models, 2,000 spectra are withheld and used as a validation set to evaluate the evolution of the models performance during the training process. The same spectra are used in the test set for each model to allow for direct comparisons of results between them, while the validation set is selected randomly at the start of each models training. We choose this split, as it allows us to get a good measure of the model performance, while maximising the size of the training set - something that is especially important given the low number of some subclassifications in the sample. The breakdown of classifications in this final sample are also shown in table \ref{tab:class_break}. Due to the small number of broad-line AGN in the final sample we do not comment on this subclass in our analysis, as there are insufficient points to draw any meaningful conclusions, though still include them in the sample for completeness.

\begin{figure*}
    \centering
    \includegraphics[width=\textwidth]{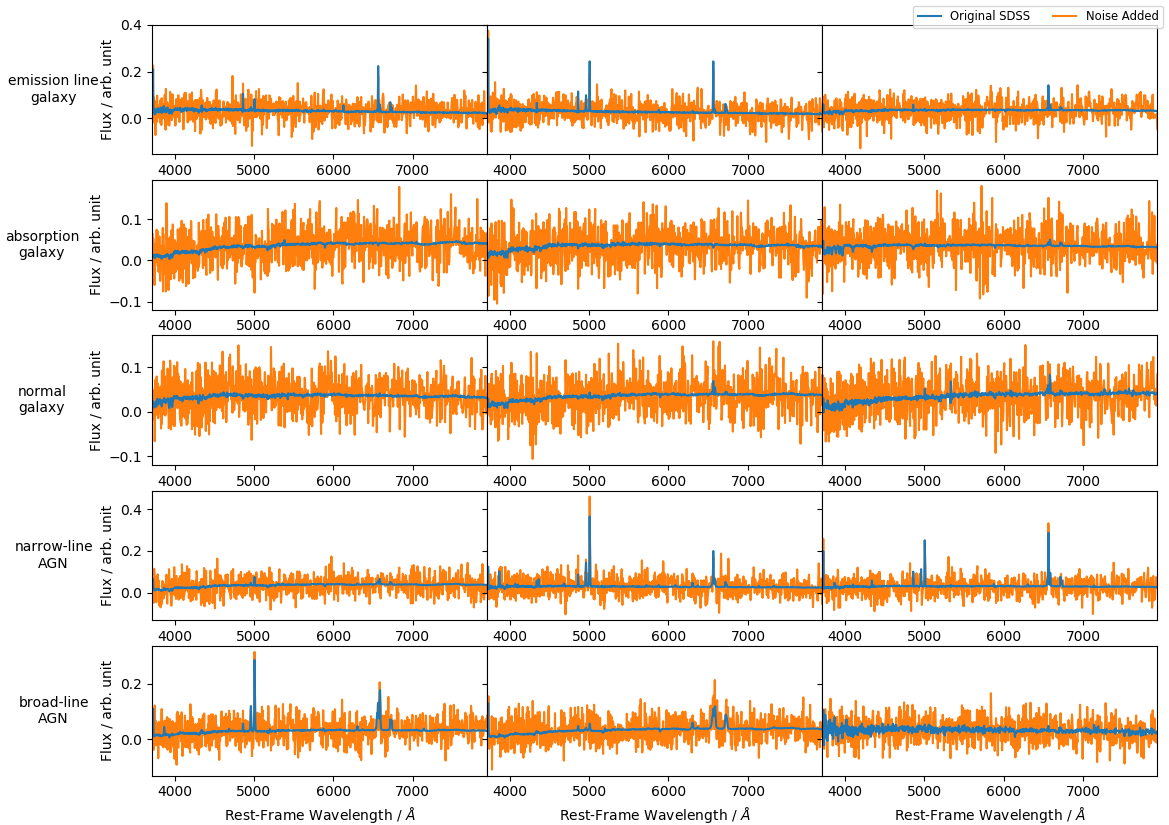}
    \caption{Example spectra from the sample comparing original SDSS (blue) and noise-added (orange) normalised flux values from various subclassifications in the sample. The spectra are processed by shifting to the rest frame and re-binning to common wavelengths. Each row corresponds to a different subclassification.}
    \label{fig:uniform_spec_ex}
\end{figure*}

Once the spectra in the sample have been selected, we pre-process them by shifting the spectra to the rest frame and re-sampling to 825 common wavelength bins, logarithmically spaced between $3700$ and $7900$~\AA, then normalise them such that the total flux is unity. This reduces the range of flux values such that the models do not need to handle inputs on varying scales. We then create noise-added spectra by injecting additional Gaussian noise. In order to determine the magnitude of this noise we look at data from the DESI survey and produce a noise model following the method of \citet{DirkNoise}. This is done by taking the errors on the observed emission lines, and interpolating to produce a continuous error model over the whole wavelength range. For simplicity however, we use a uniform noise over the whole wavelength range (rather than varying with wavelength) adopting a standard deviation of 0.04 normalised flux units, which is the upper envelope of the noise level across the DESI wavelength range. The normalisation process is then repeated on these noise-added spectra. The results in a final data set with a shape of $[n \times 825]$, where $n$ is the number of spectra in a sample/batch and 825 the number of flux values. The models we use also output data in this same shape.

For clarity, throughout this paper we refer to the spectra we use as the ground truth as the 'original SDSS spectra', and the spectra we use as inputs to the models as 'noise-added spectra'. Additionally, we refer to the spectra that are output by models as 'predicted spectra'. Examples of noise-added and original SDSS spectra from each subclassification are shown in figure \ref{fig:uniform_spec_ex}.

\section{Methods}
\label{sec:methods}
 
There have been various papers looking into using dimensional reduction tools in order to parameterise spectra \citep{Port_VAE, PAE}; however, another advantage of latent spaces these studies do not explore is the varying information content of the different dimensions. While some dimensions in a reduction will contain information on important spectral features such as emission line strengths others will mainly contain information about the noise on the data. Discarding such dimensions results in also discarding the noise they describe, while leaving the information contained within other dimensions intact. We explore two different methods of dimensional reduction as a means of this noise reduction method; the more traditional PCA method and deep learning autoencoder methods. In both cases we look at performance using 2, 4, 6, and 10 components, as in \citet{Port_VAE}.

\subsection{Principal Component Analysis}
\label{sec:methods_PCA}

Principal Component Analysis \citep[PCA, ][]{PCA_original, PCA_review} revolves around the idea of finding the principal components of a data set, and using linear combinations of them to recreate the input. These principal components are obtained by taking the covariance matrix of the data set and computing its eigenvectors and eigenvalues. These eigenvectors correspond to the principal components and the eigenvalues to the variance, and hence the information content, of the corresponding vector. Thus, by taking the eigenvectors in order of decreasing eigenvalues they are also ordered by their importance to correctly reconstruct data. For more details on the exact process see textbooks such as \citet{PCA_book_eigen}.

Inputs can then be encoded by calculating the amount of each component required to recreate them. The total number of PCA components produced is the same as the number of dimensions in the data; however, as the components are ordered by variance it is possible to use fewer components with minimal information loss by removing those at the end of the list first. Removing components in this manner results in a dimensional reduction model.

In terms of spectral data, these components take the form of their own template spectrum which control one or more features in the final spectrum. For example, one template may control the amplitude of the H$\alpha$ line in the output while another may control the strength of the {4000 \AA{}} break. In cases where a template controls multiple features they will commonly be related in some way, such as the lines in a doublet.

We apply PCA to the noise-added spectra in out training set using the \verb|sklearn| implementation \citep{scikit-learn}, producing models with 2, 4, 6 and 10 components. The test set noise-added spectra are then passed through the resulting models and their reconstruction losses evaluated using the mean square error of the result compared to the original. These values are then used as reference points for the performance of our machine learning methods.

\begin{table}
    \centering
    \begin{tabular}{c c c c}
        n component &   variance    &   \% variance & cumulative \% variance \\
        \hline\hline
           1	&    4.4e-2	&       3.2	&    3.19 \\
           2	&    9.6e-3	&       0.7	&    3.89 \\
           4	&    3.0e-3	&      0.22	&    4.50 \\
           6	&    2.8e-3	&       0.2	&    4.90 \\
          10	&    2.7e-3	&       0.2	&    5.70 \\
         677	&    1.1e-3	&     0.078	&   90.04 \\
         808	&    8.1e-4	&     0.058	&   99.04 \\
    \end{tabular}
    \caption{The variance explained by each component of the PCA analysis of the noise-added spectra. The value as a percentage of the total variance, as well as the cumulative percentage including all previous components is also shown.}
    \label{tab:PCA_noise}
\end{table}

\begin{table}
    \centering
    \begin{tabular}{c c c c}
        n component &   variance    &   \% variance & cumulative \% variance \\
        \hline\hline
           1	&    4.2e-2	&        70	&   69.77 \\
           2	&    7.7e-3	&        13	&   82.56 \\
           4	&    1.0e-3	&       1.7	&   89.82 \\
           5	&    4.8e-4	&       0.8	&   90.62 \\
           6	&    4.3e-4	&      0.71	&   91.32 \\
          10	&    1.3e-4	&      0.21	&   92.55 \\
         384	&    2.8e-6	&    4.6e-3	&   99.00 \\
    \end{tabular}
    \caption{As in Table \ref{tab:PCA_noise}, but for PCA analysis of the original SDSS spectra.}
    \label{tab:PCA_SDSS}
\end{table}

The variance for a number of the components in the PCA are given in table \ref{tab:PCA_noise}, as well as their corresponding percentage variance and the cumulative percentage variance up to that component. Of note, a total of 677 components are requires to explain 90\% of the variance, increasing to 808 for 99\%. In fact, only $\sim$6\% of the variance is explained by the first 10 components. For comparison the same information for the components when we apply PCA to the original SDSS spectra is shown in table \ref{tab:PCA_SDSS}, where it can be seen that only 5 components are required to explain 90\% of the variance, 384 to explain 99\% and the first 10 components contain nearly 93\% of the variance. This shows how the increase in overall variance of the data caused by the additional noise necessitates a larger number of components in order to adequately explain the variance, and thus accurately reproduce spectra.

\begin{figure*}
    \centering
    \includegraphics[width=\textwidth]{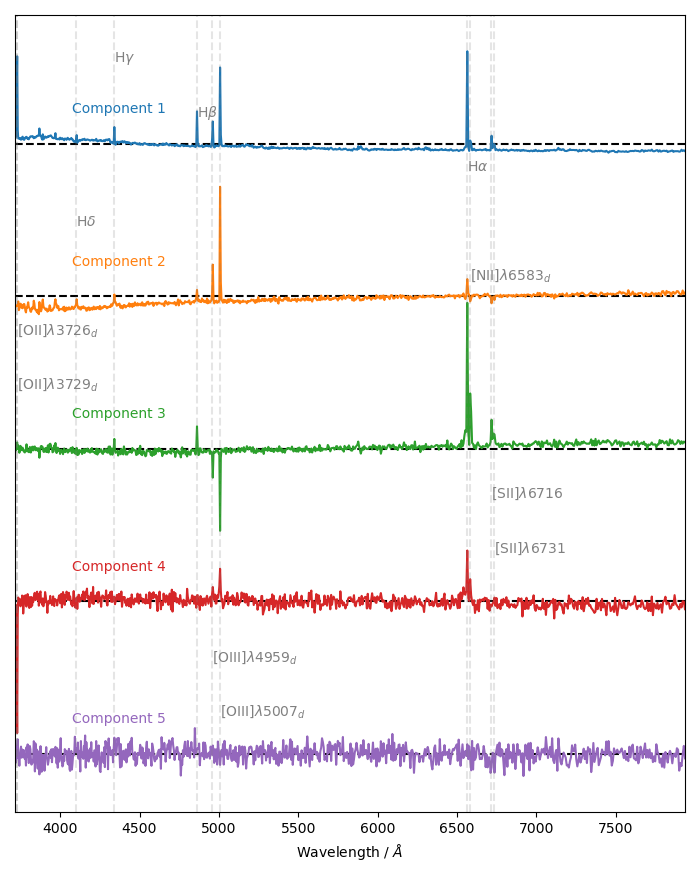}
    \caption{The first 5 components of the PCA analysis of the noise-added spectra. For each component the horizontal black dashed line corresponds to the zero point, and the vertical grey dashed lines correspond to different line features. The PCA reconstructs spectra through linear combinations of these components. Subsequent components are not included as they encode only noise.}
    \label{fig:PCA_comps}
\end{figure*}

Returning our focus to the PCA analysis of the noise-added spectra, we plot the first 5 components in figure \ref{fig:PCA_comps} in order to identify which features influence the reconstruction the most. We can see that the first component already contains the template for a large number of lines, including the Balmer series, the OII, OIII and SII doublets and NII, showing that these are an important feature when reconstructing spectra. These lines are also present in other components, such that their exact ratios are controlled through a combination of components rather than lines being controlled by individual components demonstrating that the model has picked up on relations between them. While the lines in the first component are relatively narrow, later components feature broader versions of these lines such that their widths can be controlled. Note that PCA components can have either positive or negative contributions to the predicted spectra, and so components do not have to exclusively correspond to emission or absorption.

Continuum shape are not only controlled by a superposition of components similar to line features, but are in fact also controlled by the same components that contain line features, highlighting the relation between continuum shape and lines in spectra.

The SNR of these features decreases as as we look at later components, with the 4th components containing very little information about the continuum shape. By the time the 5th component is reached continuum features have completely disappeared amongst the noise, and the small number of of line features which appear to be present, such as the SII doublet, have very low SNR. In subsequent components these line features have been lost too, such that they serve only to encode noise.

\subsection{Autoencoders}
\label{sec:methods_AE}

An autoencoder \citep[AE,][]{AE_ref} is a type of neural network which attempts to reproduce a given input by passing it through a bottleneck - a layer in the model with fewer dimensions than the input. By passing through this layer, the model is forced to learn to encode inputs into a latent space, in effect performing a dimensional reduction on the data. The structure of an AE can be split into two parts with the latent space in the middle, an encoder which handles the reduction of inputs down to this latent space, and a decoder which converts for the latent representation back out to the form of the input.

An alternative to AEs are variational autoencoders \citep[VAEs,][]{VAE_ref}, a type of AE where, rather than a single fixed value being generated for each dimension in the latent space, many normal distributions are generated instead. Values are then drawn from these distributions to use as the latent representation of the spectra. Using distributions allows for similar inputs to overlap in the latent space, leading to continuous distributions of points making interpolation between points possible. This allows for VAEs to also be used as generative models, creating new data by sampling from the latent space.

For ease of distinguishing between the two different types we henceforth refer to these two distinct architectures by their acronyms, AE and VAE, and use the full word, autoencoder, either when discussing the general concept of an autoencoder, or when referring to both architectures.

We make use of both types of model, in each case implementing a simple architecture using \verb|Keras| \citep{Keras}. The input layer takes the noise-added spectra and then passes them through a number of intermediate dense layers, which each make use of leaky ReLU activation functions. An activation function computes the value of a node in the network based on an aggregation of the values of nodes in the previous layer.

The most commonly used activation function is the rectified linear unit \citep[RELU,][]{ReLU}, which takes an input and sets it equal to zero if it is negative, and leaves positive values unchanged using the formula

\begin{equation}
    \text{ReLU}(x) = max(0, x).
\end{equation}

\noindent One problem that can arise when using this activation is that during training, a node can become stuck in a state where it always gives zero. As this is the result of the number being set to zero rather than of a calculation the model is not able to pass information to the previous layer to solve this problem and so the node 'dies', effectively reducing the number of nodes in the network which contain useful information.

Leaky ReLU is an alternative activation function which seeks to avoid this problem. It does so by modifying the function such that instead of setting negative values to zero, they are instead damped by a factor $d$ in order to still allow some information to pass through, albeit at a reduced rate \citep{leaky_ReLU}.

\begin{equation}
    \text{Leaky ReLU}(x) = max(d\cdot x, x) \quad \text{where } 0 < d < 1.
\end{equation}

\noindent In this case, we use $d = 0.3$.

During training of the model, dropout is also implemented on these layers in order to reduce overfitting, a technique by which during training the activations of each node in the network may be randomly set to zero in order to stop the model becoming over reliant on a small number of nodes \citep{dropout}.

These intermediate dense layers serve as the encoder layers, and their outputs are used to generate the latent space values, though the exact method differs for the two architectures. For the AE architecture latent space values are obtained by simply passing the output into another dense layer, called the latent layer, which produces the latent space values.

For the VAE two layers are required to produce latent space values. The first is called the parameter layer, which is also a dense layer. However, instead of directly producing the latent space values, the parameter layer generates pairs of values known as distribution parameters. Each pair corresponding to a dimension in the latent space.

The second layer in the VAE is the sampling layer. This layer takes each pair of distribution parameters and treats them as the parameters of a Gaussian distribution. Specifically, the first value in each pair represents the mean, and the second value represents the log variance of the corresponding Gaussian distribution. The sampling layer then randomly samples a value from each of these Gaussian distributions to serve as the latent space value for that dimension.

No activation functions are used for the latent layer, parameter layer or sampling layer.

\begin{figure*}
    \centering
    \includegraphics[width=\textwidth]{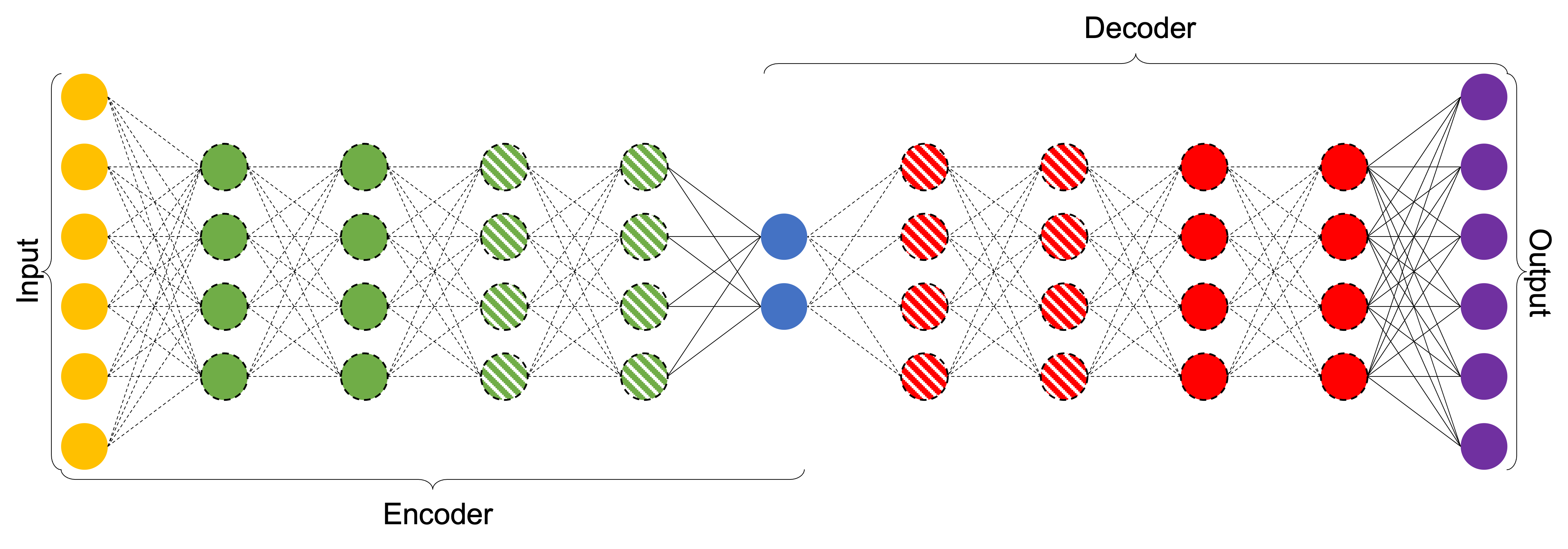}
    \caption{An example of the architecture for a 2 dimensional autoencoder model, the number of neurons in layers has been reduced for visualisation purposes. The colour of layers corresponds to their type; yellow for the input, green for the encoder intermediate layers, blue for the latent space values, red for decoder intermediate layers and purple for the output. Layers with white stripes are optional and may not be included in all models. Dashed connections correspond to layers which employ dropout during training. Neurons with no outline use no activation function and those with a dashed outline use leaky ReLU activation function.}
    \label{fig:AE_architecture}
\end{figure*}

\begin{figure}
    \centering
    \includegraphics[width=\columnwidth]{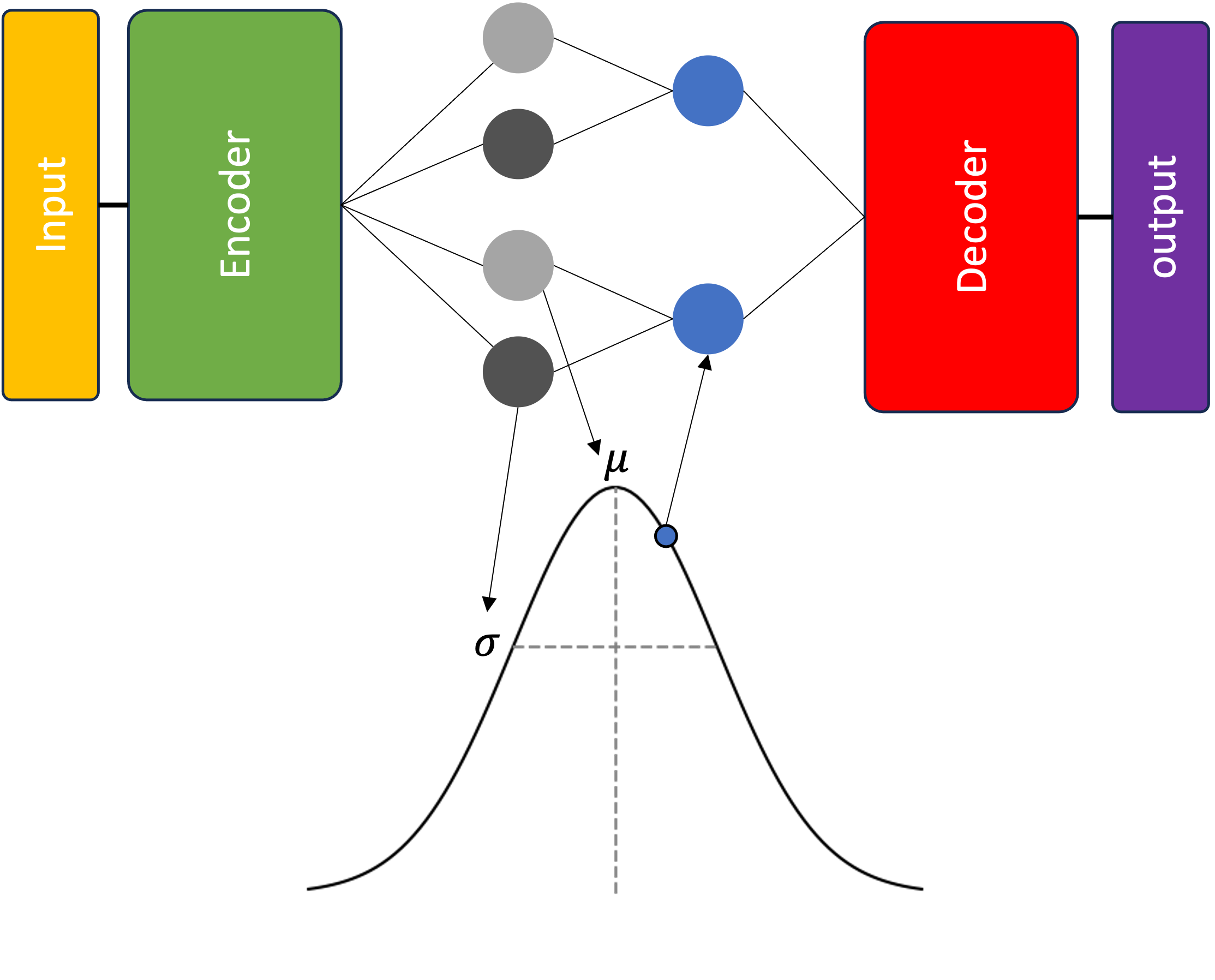}
    \caption{An example of the architecture for a 2 dimensional variational autoencoder model. The encoder and decoder architecture are largely the same as the AE (figure \ref{fig:AE_architecture}) and so repeated components been simplified here. Where the architectures differ is when the encoder produces latent space values: instead of directly generating latent space values it rather generates a pair of parameters for each latent dimension - in this case two pairs, each containing a mean, shown in light grey, and a log variance, shown in dark grey. Each pair of these parameters describes a normal probability distribution from which the latent value, in blue, is then sampled.}
    \label{fig:VAE_architecture}
\end{figure}

A simplified visualisation of the AE architecture is shown in figure \ref{fig:AE_architecture} and the VAE architecture in figure \ref{fig:VAE_architecture}.

The decoder is the same for both models; a number of densely connected intermediate layers which use leaky ReLU activation functions and employ dropout during training. The number and size of the decoder layers are the same as in the encoder, though in the reverse order. A final dense layer with no activation function then gives the model output: the predicted spectra.

These predicted spectra and the original SDSS spectra are used to measure the reconstruction loss of the models by computing the mean square error of the prediction. The reconstruction loss for a set of J spectra with K flux values is thus given by

\begin{equation}
    MSE = \frac{1}{2} \sum_{j=1}^J \sum_{k=1}^K w_{j,k} (x'_{j,k} - y_{j,k})^2,
\end{equation}

\noindent where $w_{j,k}$ is the weighting of the $k^{th}$ pixel in the $j^{th}$ spectrum, $x'_{j,k}$ the normalised flux of the pixel predicted by the algorithm and $y_{j,k}$ the normalised flux in the original SDSS spectrum. The pixel weightings are based on the inverse square of the flux error for that pixel with an additional factor of $2\times10^{-6}$ in the denominator to soften the largest weights, as in \cite{Port_VAE}.

For the AE model the reconstruction loss of the predicted spectra serves as the loss term to be optimised during training; however, for the VAE the evidence lower bound (ELBO) is optimised instead. This quantity gives a lower bound for the log-likelihood of our observed data given our model, and is used as the stochasticity introduced to our model by the latent space means we cannot calculate an exact value. The ELBO is made up of two terms,

\begin{equation}
    ELBO = L\left(y, x'\right) - D_{KL}\left(q\left(z|y\right)||p\left(z\right)\right)
\end{equation}

\noindent where L is the model likelihood, in this case the reconstruction loss, and $D_{KL}$ the KL divergence which measures the difference between the posterior of the latent values $q$ and their prior $p$. This KL term has the effect of making the distribution of inputs in the latent space match the prior, which is chosen to be the standard normal distribution as this allows for an exact analytic solution to the KL divergence to be used, as per the equation from \citet[appendix B]{VAE_ref},

\begin{equation}
    D_{KL} = \frac{1}{2}\sum_{j=1}^J(1 + log({\sigma_j}^2) - {\mu_j}^2 - {\sigma_j}^2),
\end{equation}

\noindent where ${\sigma_j}^2$ is the variance of the latent distribution for the $j^{th}$ dimension and $\mu_j$ the mean.

In both cases, optimisation is carried out using an Adam optimiser \citep{AdamOpt} with an initial learning rate of $10^{-4}$. During the training process, the optimised value is monitored such that if the value does not drop by at least $10\%$ within 5 epochs, the learning rate is decreased by a factor of 10. If this criterion is still not met within a further 5 epochs, the model training is ended early to prevent overfitting. Batch sizes of 64 are used.

The values of other hyperparameters for each model are determined through random search of 100 different models. The models are trained for up to 25 epochs. The best performing hyperparameters are then identified by comparing the mean final loss values evaluated on the validation set, then a final model is trained for 200 epochs using these hyperparameters. This process is repeated for multiple numbers of latent dimensions, each time testing 100 hyperparameter combinations. Once all of the dimensionality have been optimised their final performances are compared by taking the mean of the reconstruction loss on the test set.

Depending on the number of layers, between 4 and 6 hyperparameters are tuned for the models; the dropout rate, the number of intermediate layers in the encoder and decoder, and the number of units in each of these intermediate layers. The ranges for these parameters are:

\begin{itemize}
    \item Dropout Rate: 0 - 0.9 (steps of 0.1)
    \item Intermediate Layers: 2 - 4
    \item Layer Units: 100 - 1000 or previous layer size (steps of 100)
\end{itemize}

\noindent which are sampled uniformly.

\begin{table}
    \centering
    \begin{tabular}{c c c c c c c}
        Latent & Dropout & Intermediate & \multicolumn{4}{c}{Units}\\
        Dimensions & Rate & Layers & 1 & 2 & 3 & 4 \\
        \hline\hline
         2 & 0.8 & 4 &  900 & 800 & 400 & 300 \\
         4 & 0.5 & 4 &  900 & 900 & 500 & 400 \\
         6 & 0.7 & 4 & 1000 & 900 & 800 & 800 \\
        10 & 0.9 & 4 &  900 & 800 & 800 & 600
    \end{tabular}
    \caption{Optimal AE hyperparameters found for each number of latent dimensions in the uniform sample, found through random search.}
    \label{tab:uniform_AE_hp}
\end{table}

\begin{table}
    \centering
    \begin{tabular}{c c c c c c c}
        Latent & Dropout & Intermediate & \multicolumn{4}{c}{Units}\\
        Dimensions & Rate & Layers & 1 & 2 & 3 & 4\\
        \hline\hline
        2  & 0.8 & 4 & 1000 &  900 & 600 & 200 \\
        4  & 0.7 & 4 & 1000 & 1000 & 700 & 500 \\
        6  & 0.5 & 4 &  800 &  800 & 700 & 600 \\
        10 & 0.9 & 4 &  900 &  900 & 600 & 400
    \end{tabular}
    \caption{Optimal VAE hyperparameters found for each number of latent dimensions in the uniform sample, found through random search.}
    \label{tab:uniform_VAE_hp}
\end{table}

\begin{figure*}
    \centering
    \includegraphics[width=\textwidth]{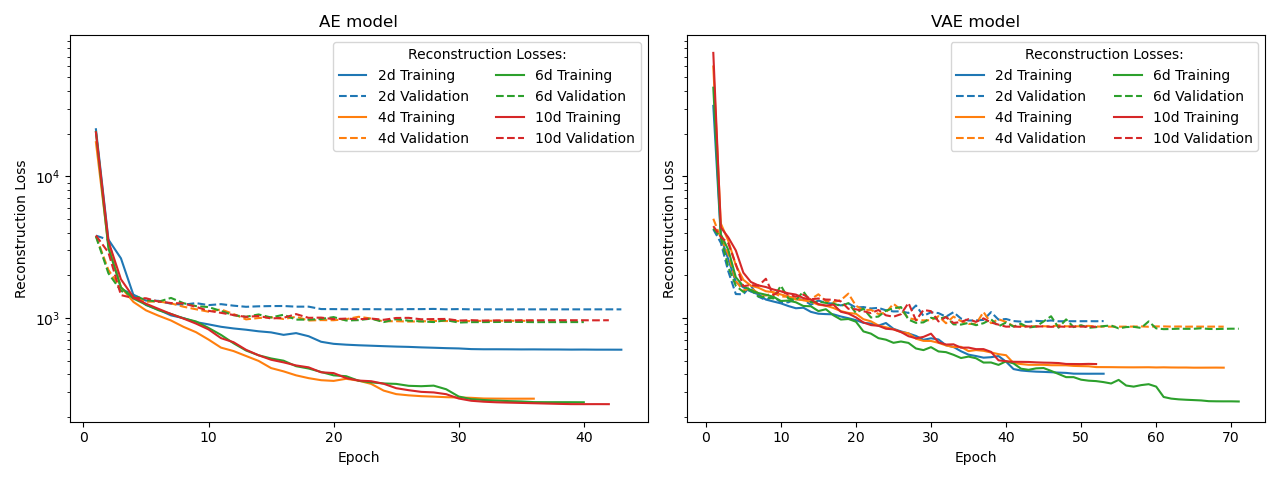}
    \caption{Reconstruction loss history during training of the autoencoder models for each of the different dimensionalities. The solid lines correspond to loss calculated using the training set after each epoch, which is shown during training, and the dashed to loss on the validation set, which the model is not shown. {\em Left:} Loss for the AE model. {\em Right:} Loss for the VAE model. Note that although training loss is lower for the AE model, validation loss is lower for the VAE model, showing that the AE model overfits the data more.}
    \label{fig:double_recon_hist}
\end{figure*}

The optimum hyperparemeter values found for each number of latent dimensions are shown in table \ref{tab:uniform_AE_hp} for the AE model, and table \ref{tab:uniform_VAE_hp} for the VAE model. The training and validation losses are also shown in figure \ref{fig:double_recon_hist}, the AE in the left panel and the VAE in the right panel.

\section{Results}
\label{sec:results}

\subsection{Clustering of galaxy classes in the latent space}

\begin{figure*}
    \centering
    \includegraphics[width=\textwidth]{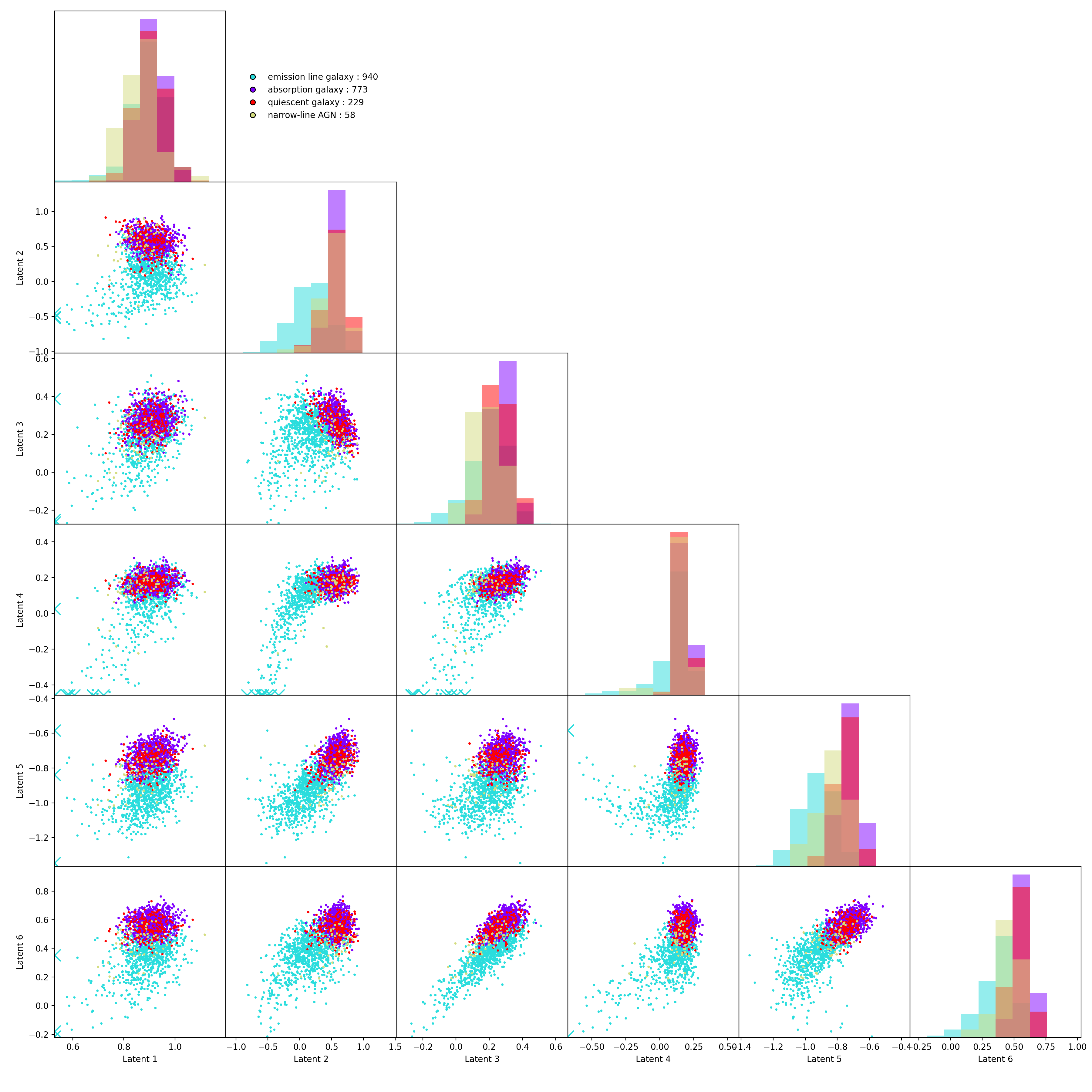}
    \caption{Latent space representation of the 6 dimensional VAE model, points are coloured based on their classification. Coloured arrows at the edge of a plot indicate points of the corresponding classification that lie outside the plotted range. Note that the order of the dimensions is arbitrary and has no relation to their information content.}
    \label{fig:latent_space}
\end{figure*}

Once we train the final models, we produce plots of their latent spaces. The latent space for the 6 dimensional VAE model is shown in figure \ref{fig:latent_space}, with points coloured based on subclassification. Note that the numbering of the latent space values is based on their order in the VAE, and as such has no significance to their importance, unlike in PCA.

Emission line galaxies, which make up the largest subclassification in the sample, are present throughout the distributions; however, the other galaxy subclassifications are much more concentrated, resulting in the latent slices taking the appearance of a main cluster with an extended tail of emission line galaxies leading off. Within this cluster, the absorption and quiescent subclasses both occupy similar spaces, making them difficult to disentangle. The narrow-line AGN appear to occupy a much more concentrated region within the cluster.

The histograms present another view of these same trends, as the emission line galaxies distribution show a tail to one side in a number of dimensions, as does the narrow-line AGN to a lesser extent. All subclasses are present in the regions where the absorption and quiescent histogram distributions overlap, with those two subclassifications overlapping almost entirely such that the model struggles to differentiate the two.

This latent space distribution of a main cluster with a tail is similar to that found by \citet{Port_VAE}; however, the roles of the different subclassifications differ. In their latent space it is instead the quiescent galaxies that form the tail, and the emission line galaxies that form the bulk of the main cluster. Note that they do not have an absorption line galaxy subclassification, instead splitting galaxies with high Balmer emission strengths between the emission line galaxies and narrow-Line AGN subclassifications based on the [NII]/H$\alpha$ line-ratio diagnostic - doing the same with our sample would result in only a single subclassification making up the main cluster as currently it is a mix of absorption line and quiescent galaxies. Despite using a different classification scheme based on the BPT diagram and H$\alpha$ equivalent width, the latent space presented by \cite{PAE} also shows overlapping classifications with regions dominated by a single class - in their case starforming and passive galaxies. Other dimensionalities of the VAE model show similar distributions, as do the AE latent spaces.
 
This makes the latent space a useful tool for identifying emission line galaxies, and also narrow-line AGN in some cases; however, the degeneracy between absorption and quiescent means that these classifications cannot be separated from one another using this method.

\subsection{Reconstruction loss and spectrum recovery}

\begin{figure}
    \centering
    \includegraphics[width=\columnwidth]{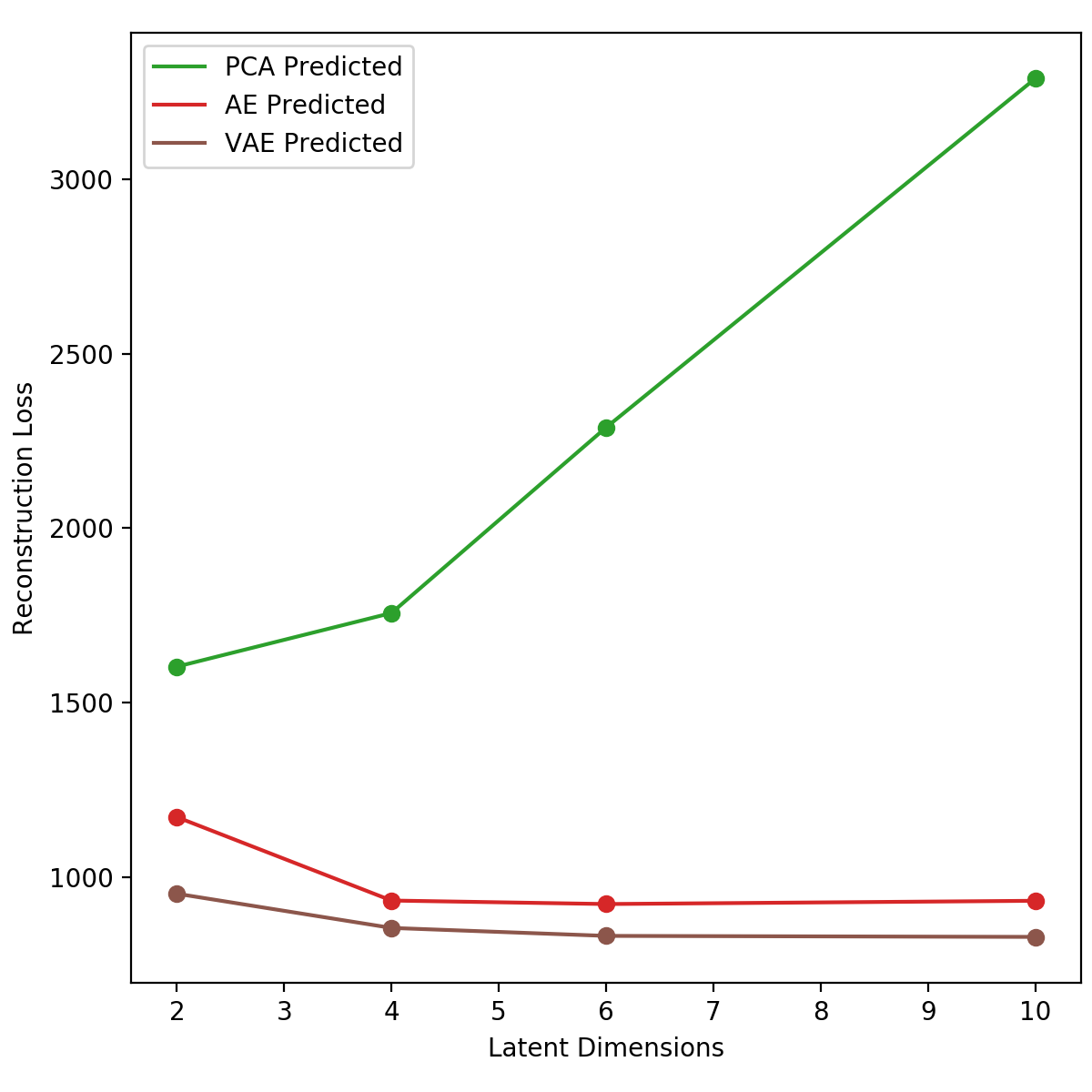}
    \caption{Comparison of how the mean reconstruction loss over the whole test set}, measured using mean squared error, of the PCA and autoencoder models vary with different numbers of latent dimensions.
    \label{fig:recon_losses}
\end{figure}

In order to analyse how well the models can reproduce the original SDSS spectra, we plot the mean reconstruction loss over the whole test set in Figure \ref{fig:recon_losses}. The figure shows that the autoencoder predictions (both from the AE and VAE) outperform the PCA model, for all dimensionalities. The PCA is most competitive in the case of the 2 dimension latent space. For the PCA, the reconstruction loss increases at a roughly constant rate as the number of dimensions increases, indicating that the first 2 dimensions contain the most information, and additional dimensions add mostly noise. Meanwhile, the two autoencoder models initially show improvement as dimensionality increases before flattening out after 6 dimensions. The largest drop in reconstruction loss for both occurs between 2 and 4 dimensions, after which point the reconstruction loss plateaus.

Despite the similarities in their behaviour however, the VAE consistently outperforms the AE for all dimensions. This result is surprising, as the AE is optimised solely on the reconstruction loss while the VAE simultaneously fits the KL divergence. However, a closer examination of the training loss shown in figure \ref{fig:double_recon_hist} reveals that the AE model has lower loss on the training data compared to the VAE.

The relative performance of the two models switches when the models are evaluated on the validation and test data, suggesting that the AE model is overfitting the training data to a higher degree, while the VAE model is able to generalise more easily. This difference can be attributed to the noise added to the latent space of the VAE during training, the randomness of which helps to reduce overfitting.

This difference in behavior between the PCA and autoencoder models as dimensionality increases can be attributed to their different approaches to dealing with noise. Gaussian noise, which adds variance to the data, can be problematic for PCA as it attempts to encode all of the variance, including that added by noise. As more dimensions are included, the model becomes increasingly overfit and unable to generalize to new data. In contrast, autoencoders can learn to filter out noise during the encoding process \citep{AE_corruption, AE_denoising}, allowing them to maintain accuracy with increasing dimensionality. This ability to ignore noise enables autoencoders to better capture meaningful features and trends in the data, whereas PCA struggles to distinguish signal from noise.

\begin{figure*}
    \centering
    \includegraphics[width=\textwidth]{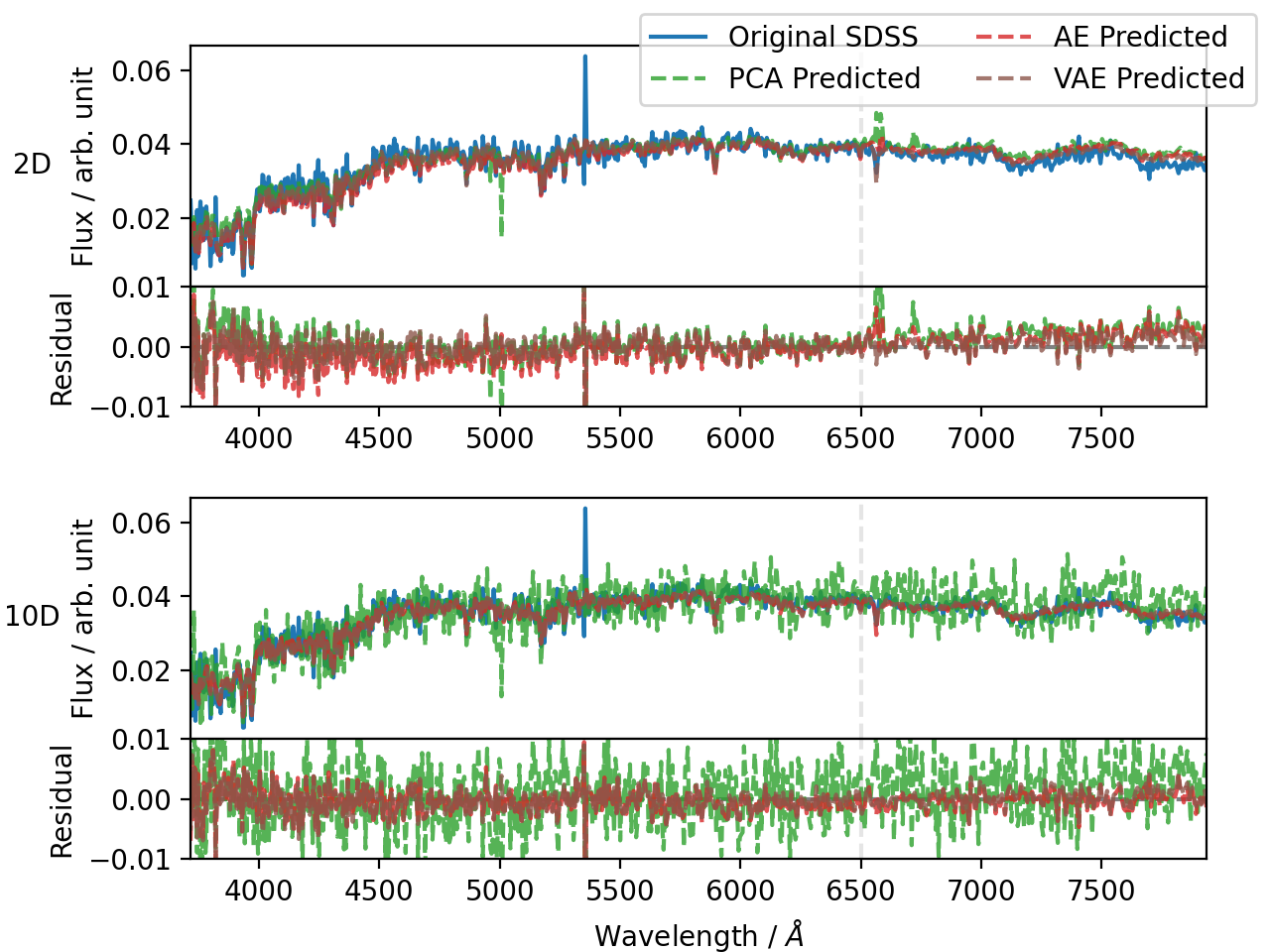}
    \caption{Comparison of the model reconstructions of an absorption galaxy from the test set. The top panel shows the 2 dimensional model reconstructions and the bottom the 10 dimensional models. The residuals for each plot are presented in separate panels beneath their respective reconstructions. The vertical dashed line corresponds to 6500 \AA, beyond which the continuum is commonly over/under predicted for lower dimension models.}
    \label{fig:recon_comp}
\end{figure*}

To visualise how these losses translate into spectral reconstructions we plot an example spectrum from the sample (from an absorption galaxy) in the top panel of Figure \ref{fig:recon_comp}, overlaying the 2 dimensional predicted spectra from each of the methods. The 2 dimensional model is chosen, as this is the dimension for which the performance of the PCA and autoencoder models are most similar. From this, we can see that the position of absorption and emission features are reproduced well, as is the galaxy continuum up to $\sim6500$\AA{}, after which point it begins to over predict the continuum flux. We visually inspected many spectra and their reconstruction, and notice that this difficulty of the 2 dimensional PCA and AE models to reproduce the continuum at the longest wavelengths is common across the sample. 

These discrepancies are likely a result of the fact that the information content of optical spectra is largest at shorter wavelengths \citep[e.g. around the 4000\AA{} break, see][]{ferreras22}, and so greater improvements in the reconstruction loss are achieved by correctly fitting these more information-rich spectral regions. Thus, errors in reproducing the spectra at $\lambda > 6500$\AA{} are less costly, and so the models do not learn to handle them as effectively. 

Looking instead at higher dimensionality reconstructions, e.g. 10 dimensions as shown in the bottom panel of figure \ref{fig:recon_comp}, it is clearly visible that the additional parameters available allow the AE and VAE models to better reproduce the full spectrum (as evidenced by the lower reconstruction loss). Meanwhile, the PCA reproduction significantly increased noise in the continuum, further increasing the previous conclusion that additional dimensions mostly encode noise.   

In order to test what spectra the models are worst at reproducing, we identify the predicted spectra with the highest reconstruction losses for each model. We find that the spectra with the greatest loss is the same across all models. Additionally, looking at the top 5 highest reconstruction losses there is a large amount of overlap in the spectra that occur between all of the models.

Visually inspecting the original SDSS versions of these common spectra, while some of them have low SNR, others have SNR comparable to spectra which are well reproduced. Instead, what is common between the group is an unusual feature where the initial few flux values of the spectra (up to $\sim$3725\AA) are all significantly higher than the continuum, being instead comparable to the peak H$\alpha$ flux values seen in emission line galaxies. While this feature does overlap with the [OII]3726\AA{} doublet, it is too flat to be these lines - though in some cases it does appear the line is superimposed on top of this feature - and so we believe this not to be a physical feature but an error in the original SDSS spectra.

In the predicted spectra this feature is not present, and is instead replaced with an OII emission line, and the continuum values are consistent with those at higher wavelengths.

The fact that the presence of this unphysical feature causes spectra to be badly reproduced shows that these models, while agnostic to the physics responsible for spectral features, still do not blindly reproduce unphysical features. Instead in the presence of anomalies the models attempt to reconcile the errors with what they saw during training.

\subsection{Line flux recovery}
\label{sec:lineFlux}

\begin{figure*}
    \centering
    \includegraphics[width=\textwidth]{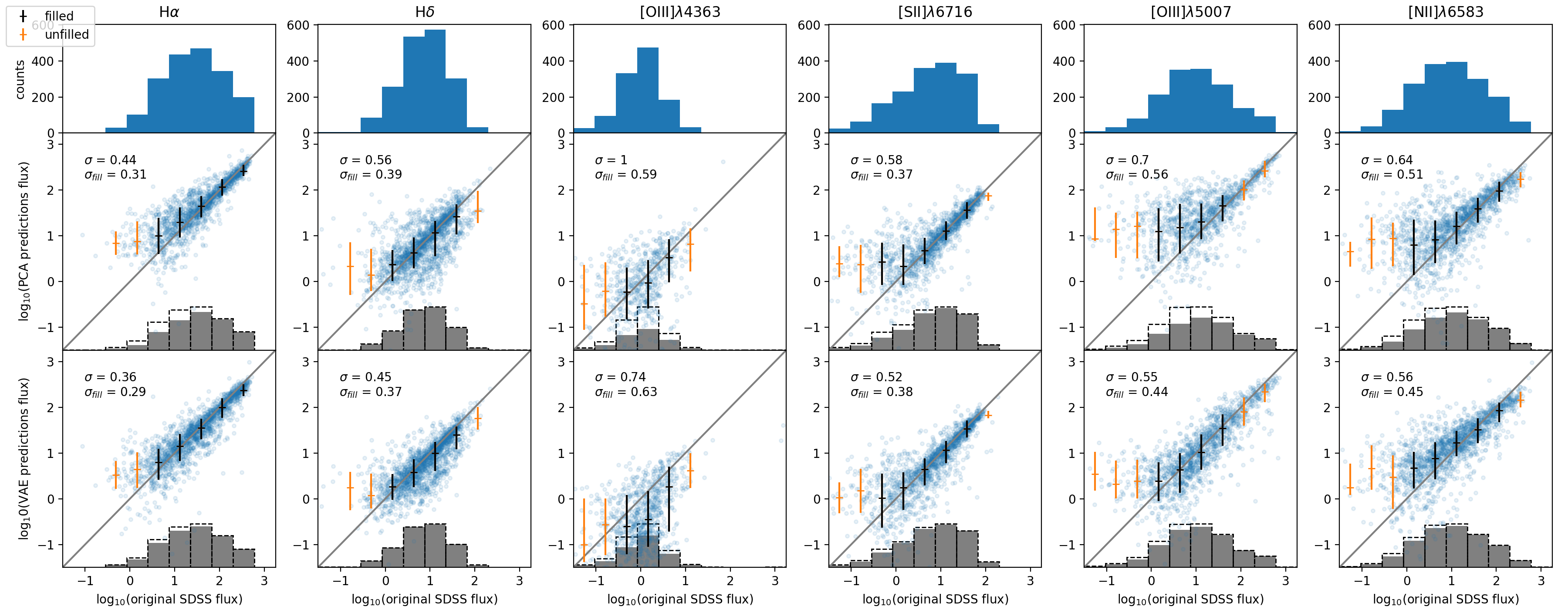}
    \caption{Line peak flux comparison between the original SDSS spectra and the different 4 dimensional model predictions (rows) for various lines (columns). The AE method is not shown, though gives similar results to the VAE method. Data are split into 10 bins based on the original SDSS flux value. The error bars show the median flux for each bin with errors corresponding to the 16th and 84th percentiles. The standard deviation for all points is shown in each panel, as is the standard deviation for points in full bins only. Full bins are defined as those containing more than $N/(n_{bins} + 1)$ points and shown in black, whereas unfilled bins are shown in orange. The histograms in the top row show the distribution of the original SDSS flux value for each line. In subsequent rows this is reproduced as a dashed outline, with the grey bars showing how many lines in each bin were successfully fit using pPXF.}
    \label{fig:line_peak_flux}
\end{figure*}

\begin{table*}
    \centering
    \begin{tabular}{c c c c c c c c c c c c c}
        & \multicolumn{2}{c}{H$\alpha$}&\multicolumn{2}{c}{H$\delta$}&\multicolumn{2}{c}{OIII 4363}&\multicolumn{2}{c}{SII 6716}&\multicolumn{2}{c}{OIII 5007}&\multicolumn{2}{c}{NII 6583} \\
        Model & $\sigma$ & $\sigma_{{fill}}$ & $\sigma$ & $\sigma_{{fill}}$ & $\sigma$ & $\sigma_{{fill}}$ & $\sigma$ & $\sigma_{{fill}}$ & $\sigma$ & $\sigma_{{fill}}$ & $\sigma$ & $\sigma_{{fill}}$ \\
        \hline\hline
        2 dimensions \\
        PCA	&	0.52	&	0.33	&	0.51	&	0.39	&	0.82	&	0.61	&	0.64	&	0.44	&	0.65	&	0.54	&	0.68	&	0.50 \\
        AE	&	0.45	&	0.36	&	0.52	&	0.35	&	0.75	&	0.60	&	0.61	&	0.45	&	0.53	&	0.42	&	0.64	&	0.48 \\
        VAE	&	0.42	&	0.30	&	0.56	&	0.37	&	0.75	&	0.62	&	0.53	&	0.41	&	0.49	&	0.38	&	0.59	&	0.49 \\
        \hline
        4 dimensions \\
        PCA	&	0.44	&	0.31	&	0.56	&	0.39	&	1.00	&	0.59	&	0.58	&	0.37	&	0.70	&	0.56	&	0.64	&	0.51 \\
        AE	&	0.41	&	0.34	&	0.54	&	0.34	&	0.71	&	0.59	&	0.55	&	0.42	&	0.52	&	0.41	&	0.62	&	0.52 \\
        VAE	&	0.36	&	0.29	&	0.45	&	0.37	&	0.74	&	0.63	&	0.52	&	0.38	&	0.55	&	0.44	&	0.56	&	0.45 \\
        \hline
        6 dimensions \\
        PCA	&	0.43	&	0.37	&	0.70	&	0.42	&	0.82	&	0.55	&	0.58	&	0.38	&	0.72	&	0.57	&	0.62	&	0.49 \\
        AE	&	0.43	&	0.35	&	0.52	&	0.36	&	0.72	&	0.57	&	0.51	&	0.39	&	0.52	&	0.44	&	0.61	&	0.48 \\
        VAE	&	0.37	&	0.29	&	0.41	&	0.33	&	0.76	&	0.61	&	0.53	&	0.42	&	0.54	&	0.43	&	0.59	&	0.48 \\
        \hline
        10 dimensions \\
        PCA	&	0.55	&	0.37	&	0.64	&	0.40	&	0.86	&	0.61	&	0.60	&	0.42	&	0.73	&	0.56	&	0.71	&	0.57 \\
        AE	&	0.41	&	0.32	&	0.48	&	0.38	&	0.80	&	0.60	&	0.59	&	0.40	&	0.67	&	0.45	&	0.60	&	0.49 \\
        VAE	&	0.43	&	0.31	&	0.45	&	0.36	&	0.77	&	0.60	&	0.53	&	0.38	&	0.64	&	0.43	&	0.59	&	0.48 \\
        \hline
    \end{tabular}
    \caption{The standard deviations for the peak line flux predictions from our models vs the original SDSS spectra for a variety of lines. For each line, we show both the standard deviation over the full range of line fluxes as well the standard deviation calculated using only full bins - defined as those containing more than $N/(n_{bins} + 1)$. points}
    \label{tab:line_peak_flux}
\end{table*}

While we have shown in the previous section that the overall spectral shape and features are recovered by the autoencoder models, we now need to quantify the accuracy of the emission line fluxes that can be recovered. We use pPXF \citep{cappellari04,cappellari17, ppxf} to fit the stellar continuum and emission line fluxes of both the the original SDSS spectra and the models (both the PCA and VAE). The two sets of fluxes are plotted against each other in Figure \ref{fig:line_peak_flux}; this allows us to asses how well the line fluxes recovered from the noise-added spectra by the VAE (or PCA) compare to those directly measured from the original SDSS spectra. We show the results for the 4 dimensional model.  

We evaluate the performance of the models both using the full range of fluxes probed, and only on the well sampled regions (the latter labelled as "filled bins" in Fig. \ref{fig:line_peak_flux}). We define such "filled bins" as those where the number of measurements is $> N/(n_{bins} + 1)$, where $N$ is the total number of points and $n_{bins}$ the number of bins used. We use this criteria as it selects bins with at least as many spectra in them as we would expect from a uniform distribution. Figure \ref{fig:line_peak_flux} includes the scatter of the fitted values over both ranges in each panel. These values are shown for all dimensionalities, as well as for the AE models, in Table \ref{tab:line_peak_flux}.

The fluxes recovered by the VAE compare very well with the SDSS reference over the flux range that is well sampled. Over this interval, the scatter is 0.3 dex for H$\alpha$, the line typically with the highest S/N, increasing to 0.45 dex for fainter lines such as [NII]$\lambda$6583 and [OIII]$\lambda$5007.  For all the emission lines studies, the flux values recovered by the VAE and PCA are overestimated at the lowest input flux values, and the scatter increases. 
The overestimation of the fluxes in the faint line regime is more severe for the PCA predictions than for the VAE method for the H$\alpha$, [OIII]$\lambda$5007 and [NII]$\lambda$6583. The overall scatter of the fluxes recovered by the PCA method is also higher overall, and it is more common for line flux values not to be recovered at all in the PCA predictions.

Comparing the scatter of VAE fluxes to the scatter obtained if we treat the uncertainty on fitted SDSS flux values as the error, we find the model scatter to be a factor $\sim2$ greater for all lines with the difference being smallest for H$\delta$ and greatest for [OIII]$\lambda$5007.

This shows that the flux values obtained after applying our model to noise-added spectra have comparable accuracy to those obtained from the original SDSS spectra and that through the application of our model, data with lower SNR can be included in scientific analysis with minimal loss of accuracy when compared to higher SNR data. This accuracy loss can be further reduced if we are able to avoid the areas of systematic over prediction in our model. One way to do this is by applying appropriate flux cuts to keep line fluxes above these regions.

\section{Test case: the mass-metallicity relation}
\label{sec:metallicity}

To further assess the reliability of autoencoder-predicted emission line fluxes from noisy optical spectra for galaxy evolution studies, we investigate how well they can be applied to the study of the mass-metallicity relation. We choose metallicity for this test, as they require accurate measurements of several emission lines. 

We create a new sample for this test, consisting of only star-forming galaxies, which is where metallicity can be accurately calculated, with sufficient S/N in the emission lines without contamination from AGN activity. This sample is selected using the criteria listed in Section \ref{sec:data}, with the additional condition that the galaxies belong to the star-forming subclass in the MPA-JHU DR7 catalogue, which is based on PCA analysis by David Schlegel. This reduces the number of galaxies to 43,952.

These selected galaxies cover a narrower range of H$\alpha$ flux values compared to the previously selection. Thus when binning them based on H$\alpha$ flux we use a reduced range of ~$10^{1.4}-10^{3.2}$ 1e-17 erg/s/cm$^2$, still using 10 logarithmically spaced bins. However, we note that the highest luminosity bin is not completely filled, containing only 778 spectra. As a result we are unable to select a full 1,000 spectra from it, causing this sample to be slightly smaller than the previous one. The other bins contain sufficient spectra to fill them. The final sample contains 9,778 spectra, with 1,956 used for testing and 7,882 for training. Of the training set 1,956 are reserved for validation. This follows the same proportions as used previously.

To make this test as relevant as possible for future work, we switch from using simple uniform noise to using the full noise model based on the performance of the DESI survey (see section \ref{sec:data}). We also scale the input low-redshift, high-S/N SDSS data to simulate how the galaxies would appear in the DESI survey if they were all at $z=0.1$. To help differentiate between spectra with noise added using this method and those using the method discussed in section \ref{sec:data}, we refer to these spectra as 'DESI-like spectra'.

The VAE model is then re-trained on these DESI-like spectra, and the model applied to the rest of the spectra to perform our test. Line fluxes are once again measured using pPXF with the gas-phase metallicities then calculated using the calibration of \cite{PP04}:

\begin{equation}
    12+\log({\rm O/H})=8.73-0.32 \times \log \left( \frac{[{\rm O III}] \lambda5007/{\rm H}\beta}{[{\rm N II}] \lambda6584/{\rm H}\alpha} \right). 
    \label{eq:PP04}
\end{equation}

\noindent The median error on metallicity is ~0.09 for each of the three different types of spectra we investigate.

\begin{figure*}
    \centering
    \includegraphics[width=\textwidth]{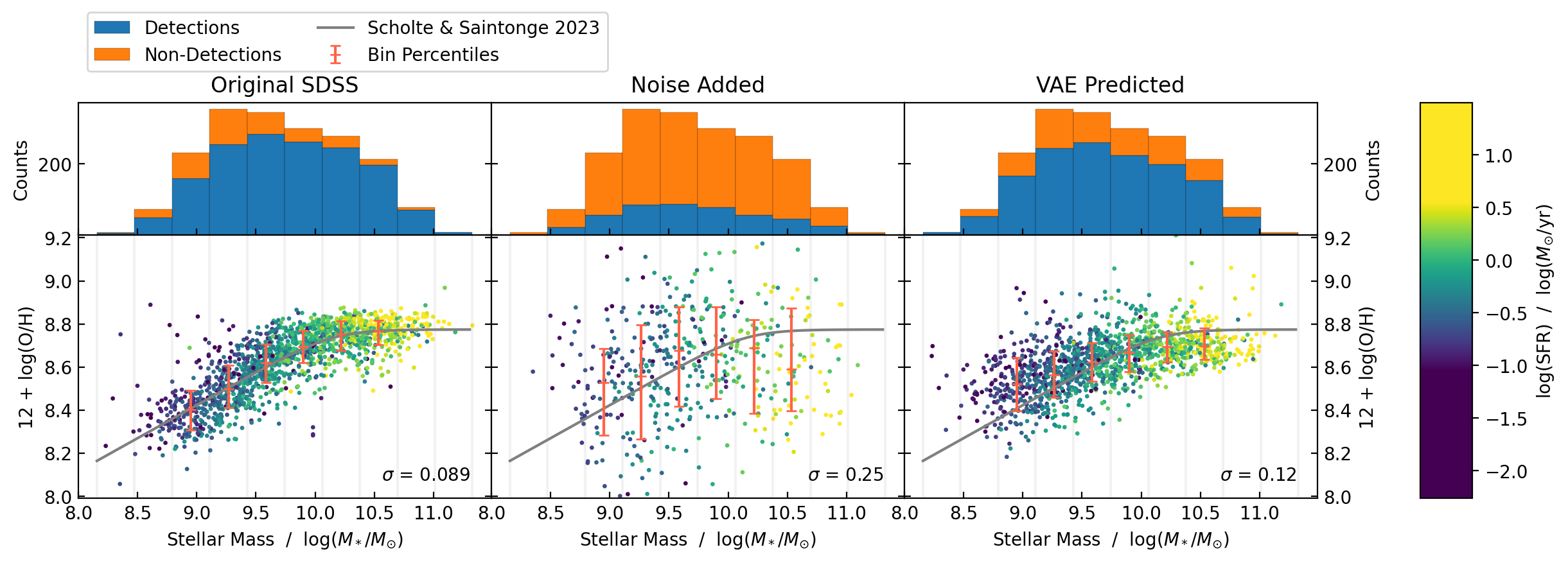}
    \caption{The mass-metallicity relation (MZR) for a sample of star-forming galaxies. {\em Left:} the MZR relation from the original SDSS sample ($z<0.05$).  {\em Middle:} The MZR for DESI-like spectra - the same galaxies, but transformed to mimic the S/N of DESI spectra at $z=0.1$.  {\em Right:} the MZR of the predicted spectra when the DESI-like spectra are passed through the 6 dimensional VAE model.  In all panels, the mass distribution of the full sample (orange histogram) is compared to that of the galaxies with sufficient signal to enable a metallicity measurement (H$\alpha$ S/N$>15$, blue histogram). The points are coloured based on SFR, and in each panel the standard deviation around the MZR of \citet{DirkNoise} is given. For sufficiently filled bins error bars are plotted show 16th, 50th and 86th percentile of the metallicity. The median error on metallicity is ~0.09 for each of the three different types of spectra.}
    \label{fig:MZR}
\end{figure*}

We show the mass-metallicity relation (MZR) of the test set in Figure \ref{fig:MZR}. The left panel is the reference: it is the MZR derived from the original SDSS spectra of galaxies at $z<0.05$. The middle panel is the MZR as would be recovered by directly measuring the emission line fluxes of the DESI-like spectra. Not only is the shape and scatter of the MZR very poorly recovered, but only a small fraction of the galaxies have high-enough signal in the emission lines to make a metallicity measurement possible \citep[we require S/N$>15$ in the H$\alpha$ line, following][]{DirkNoise}. Finally, the right-hand panel shows the MZR using the line fluxes from the predicted spectra - when the DESI-like spectra are passes through the 6 dimensional VAE model.

Applying the VAE model to the DESI-like spectra both increases the number of galaxies with measurable metallicites back to similar levels as in the original SDSS spectra, and allows for both the shape and scatter of the MZR to be recovered. The slightly increased scatter (0.12 dex as opposed to 0.09 in the original SDSS sample) mostly comes from the low mass / low metallicity regime, where we have seen in Fig. \ref{fig:line_peak_flux} that line fluxes can be overestimated by the VAE model. 

There is significant interest in studying not just the shape and redshift evolution of the MZR, but also the nature of its scatter, including any dependence on third parameters.  The consensus is that the scatter in the MZR can be best explained by variations in cold gas contents \citep[e.g.][]{hughes13,bothwell16,brown18,chen22,DirkNoise}, which also manifest themselves as a correlation between MZR scatter and the star formation rate \citep[e.g.][]{ellison08, lara-lopez10, mannucci10}. To show that the VAE-recovered line fluxes can be used to recover not just the shape and total scatter of the MZR, but also the systematic trends within the scatter, the points in Fig. \ref{fig:MZR} are color-coded by total SFR. Just like in the original SDSS sample (left panel), the VAE predicted measurements (right panel) show the trend that at fixed stellar mass the lower metallicity galaxies have lower SFRs (and vice versa). 

This proof-of-concept test shows that spectral denoising using VAEs makes it possible to retrieve accurate line fluxes and their derived quantities for galaxy spectra which otherwise would not be usable directly.

\section{Conclusions}
\label{sec:conclusions}

We used autoencoders to de-noise galaxy spectra, comparing to the performance of a more traditional PCA method. We use a sample of SDSS spectra, selected to have a uniform distribution of H$\alpha$ line fluxes. The autoencoder and PCA models are trained on noise-added spectra, using the original SDSS spectra as ground truth when assessing the performance of the models.  

Training models with varying numbers of latent dimensions, we find little variation in the latent spaces produced. In each case, emission line galaxies separate themselves out from other classifications, whereas narrow-line AGN are the most clustered.

The reconstruction loss of the PCA model increases with dimensionality, showing that most of the information is encoded in the first 2 dimensions. Meanwhile, the autoencoder models outperform the PCA method in all number of latent space dimensions tested. The variational autoencoder outperforms the autoencoder for all dimensions, but both show significantly smaller reconstruction loss than the PCA.

For low dimensional reconstructions, the continuum is frequently over/under predicted in the higher wavelength range of our spectra ($>6500$\AA). This is due to the lower information content in these regions relatively to lower wavelengths \citep{ferreras22}, meaning that the limited number of features which can be encoded by low dimension representations do not include those more information-poor spectra regions. Once more dimensions are added to the model such that more features can be encoded these over/under predictions become less common.

We fit the original SDSS and predicted spectra with pPXF, and compare the measured emission line fluxes. The VAE model reproduces the original SDSS values well over the range of fluxes well sampled in the training set, with standard deviations of 0.3-0.5 dex depending on the spectral line - a factor 2 greater than those obtained from fits to the original SDSS spectra.

To demonstrate the validity of the emission line fluxes obtained from VAE predicted spectra, we apply the VAE model to a sample of star-forming galaxies with added noise consistent with the DESI survey at $z=0.1$. We reproduce the shape and scatter of the original MZR and show that dependency of the scatter on a third parameter is preserved.

These results show the potential for machine learning techniques as a part of data analysis pipelines, as a method of maximising the data retrieved from spectra. While a loss of accuracy when working with low SNR data is inevitable we find that by using autoencoder models this loss can be reduced to enable for more accurate scientific analysis. Further work on these techniques, such as by expanding training data into currently under dense regions in order to improve the model's prediction capabilities within them, can reduce this loss of accuracy even further.

Reducing the SNR requirements in order to perform emission line analysis on spectra allows for our understanding of galaxies to be pushed forward both directly, through allowing analysis of trends in higher redshifts objects previously only observable in closer, higher SNR observations, and indirectly by allowing for more data to be gathered during surveys. Reductions in the required SNR of spectra in surveys means less time is required observing an object in order to obtain useable data, increase the rate at which targets can be observed. It also allows for the redshift range over which surveys are able to gather usable data to be expanded as they are able to observe regions where data were previously too noisy without the need for modifications to the instruments themselves.

%%%%%%%%%%%%%%%%%%%%%%%%%%%%%%%%%%%%%%%%%%%%%%%%%%

\section*{Acknowledgements}

SV acknowledges support from the European Research Council (ERC) under the European Union’s Horizon 2020 research and innovation programme MOPPEX 833460.

In this work, we make use of the following python packages: NumPy \citep{numpy}, SciPy \citep{scipy}, pandas \citep{pandas}, tqdm \citep{tqdm}, Astropy \citep{astropy:2013, astropy:2018, astropy:2022}, astroML \citep{astroML}, ppxf \citep{cappellari04,cappellari17, ppxf}, scikit-learn \citep{scikit-learn}, TensorFlow \citep{tensorflow}, Keras \citep{Keras}, KerasTuner \citep{kerastuner} and matplotlib \citep{matplotlib}.

\section*{Data Availability}
 
In this work we make use of the MPA-JHU catalogues which contain emission line analysis for SDSS DR7, in addition to collating information from other sources. These catalogues are available at \url{https://wwwmpa.mpa-garching.mpg.de/SDSS/DR7/}.


\bibliographystyle{mnras}
\bibliography{example} % if your bibtex file is called example.bib

%%%%%%%%%%%%%%%%% APPENDICES %%%%%%%%%%%%%%%%%%%%%

%\appendix


%If you want to present additional material which would interrupt the flow of the main paper, it can be placed in an Appendix which appears after the list of references.

%%%%%%%%%%%%%%%%%%%%%%%%%%%%%%%%%%%%%%%%%%%%%%%%%%

% Don't change these lines
\bsp	% typesetting comment
\label{lastpage}
\end{document}